\documentstyle[12pt,preprint,tighten,aps]{revtex}
\begin{document}
\draft
\preprint{\begin{tabular}{l}
Imperial/TP/94--95/45
\\
hep-th/9506187
\end{tabular}}

\newcommand{\bb}{\begin{equation}}
\newcommand{\ee}{\end{equation}}
\newcommand{\w}{{\mbox{\tiny $\wedge$}}}
\newcommand{\pp}{\partial}

\title{The local degrees of freedom of higher
dimensional pure Chern-Simons theories}

\author{M\'aximo Ba\~nados$^{a,b}$,
Luis J. Garay$^{b}$ and Marc Henneaux$^{a,c}$}

\address{$^{(a)}$ Centro de Estudios
Cient\'{\i}ficos de Santiago, Casilla 16443, Santiago, Chile \\
$^{(b)}$ Theoretical Physics Group, The Blackett Laboratory,
Imperial College,  London SW7 2BZ, UK  \\
$^{(c)}$ Universit\'e
Libre de Bruxelles, Campus Plaine, C.P. 231, B-1050, Bruxelles,
Belgium. }

\date{27 June 1995}

\maketitle

\begin{abstract}

The canonical structure of higher dimensional pure Chern-Simons
theories is analysed.  It is shown that these theories have
generically a non-vanishing number of local  degrees of freedom,
even though they are obtained by means of a topological
construction.  This number of local degrees of freedom is
computed as a function of the spacetime dimension and the
dimension of the gauge group.

\end{abstract}

Three-dimensional pure Chern-Simons theory is well known to
possess higher dimensional generalizations.   These
generalisations are theories in $2n+1$
dimensions  constructed from characteristic
classes in $2n+2$ dimensions in exactly the
same way as three-dimesional Chern-Simons theory is built out of
the four-dimensional characteristic classes.  More precisely, if
$F^a$ is the curvature 2-form $F^a = dA^a + \frac{1}{2}
f^a_{bc} A^b \w A^c$
associated to the gauge field 1-form $A^a$, where $f^a_{bc}$ are
the structure constants of the gauge group, and $g_{a_1\ldots
a_{n+1}}$ is a rank $n+1$, symmetric tensor invariant under the
adjoint action  of the gauge group, then one defines the
Chern-Simons Lagrangian ${\cal L}_{CS}^{2n+1}$ through the
formula
\bb
d{\cal L}_{CS}^{2n+1}=g_{a_1\ldots a_{n+1}} F^{a_1}\w\cdots\w
F^{a_{n+1}}.
\label{l2n+1}
\ee
The three-dimensional case is obtained by taking $n=1$, which
yields
$
d{\cal L}^3_{CS}=g_{ab} F^a\w F^b,
$
where $g_{ab}$ is an invariant metric on the Lie algebra
(necessarily proportional to the Killing metric if the Lie
algebra is semisimple).

The Chern-Simons action $I=\int_M {\cal L}_{CS}^{2n+1}$ is
invariant under standard gauge transformations
\bb
\delta_\epsilon A^a_\mu = D_\mu \epsilon^a.
\label{gauge}
\ee
It is also invariant under spacetime diffeomorphisms,
$\delta_\eta A^a_\mu = \pounds_\eta A^a_\mu$, since ${\cal
L}_{CS}^{2n+1}$ is a ($2n+1$)-form.  The spacetime
diffeomorphisms can also be represented by
\bb
\delta_\eta A^a_\mu = \eta^\nu F^a_{\mu \nu}.
\label{diff2}
\ee
Indeed, these symmetries differ from the Lie derivative only by
a gauge transformation and are often called improved
diffeomorphisms \cite{Jackiw}.  If the only symmetries of the
Chern-Simons action are the diffeomorphisms (\ref{diff2}) and
the gauge transformations (\ref{gauge}), then we shall say that
there is no accidental gauge symmetry.  How this translates into
an algebraic condition on $g_{a_1\ldots a_{n+1}}$ will be
described precisely below.

Of particular interest are the Chern-Simons theories with gauge
group $SO(2n+1,1)$ or $SO(2n,2)$ in $2n+1$ dimensions because
they define gravitational theories \cite{Chamseddine}.  For
$n=1$, one recovers the standard Chern-Simons formulation of
Einstein gravity with a cosmological constant \cite{at86}.  For
$n>1$, one gets the Einstein-Hilbert action supplemented by
Lovelock terms \cite{Lovelock} with definite coefficients.
These gravitational theories admit intriguing black
hole solutions \cite{BTZ1} generalizing the three-dimensional
black holes of Ref. \cite{BTZ2}.

One of the striking features of Chern-Simons theory in three
dimensions is the fact that it has no local degrees of freedom.
This is because the equations of motion
\bb
g_{aa_1\ldots a_{n}} F^{a_1}\w\cdots\w F^{a_{n}} = 0
\label{equmotion}
\ee
reduce to $F^a = 0$ in the three-dimensional case.  Thus, the
space of solutions of  Chern-Simons theory in three dimensions
is the finite-dimensional moduli space of flat connections
modulo gauge transformations. [Note that the diffeomorphisms
lead to no further quotientizing because they vanish on-shell].

Since the higher dimensional Chern-Simons theories are
constructed along the same topological pattern as their
three-dimensional analog, one may wonder whether they are also
devoid of local excitations and have only global degrees of
freedom.  One of the purposes of this letter is to explain why
this is not the case.  We also count explicitly the number of
local degrees of freedom as a function of the dimensions of
spacetime and of the gauge group.  It turns out that the crucial
ingredient that controls the whole analysis  is
the invariant tensor $g_{a_1\ldots a_{n+1}}$.

We start the discussion with the five dimensional case ($n=2$)
and an $N$-dimensional abelian group ($G= U(1)^N$).  This case
already contains  all the main points that we want to address
and is particularly simple because the invariance condition
imposes no restriction on the tensor $g_{a_1\ldots a_{n+1}}$.
We shall deal with the general situation of an arbitrary gauge
group afterwards.

Assume first that there is only one single abelian field.  The
equations of motion imply $F \w F = 0$, i.e. $F$ has at most
rank $2$.  In the generic case, $F$  has exactly rank 2 (in the
space of solutions of $F \w F = 0$, the solution $F = 0$ has
measure zero).
Since $F$ is a closed $2$-form, one may bring it locally to the
canonical form $F = dx^1 \w dx^2$ by a diffeomorphism (Darboux
theorem for presymplectic forms of rank 2).  Thus, the quotient
space of the solutions of the equations of motion modulo the
gauge transformations (\ref{gauge}) and spacetime
diffeomorphisms (\ref{diff2}) has locally one and only one
solution.  This implies that the theory has no local degrees of
freedom, in agreement with the findings of Ref.
\cite{fp89}.

The case of a single abelian gauge field is, however, a poor
representative of what happens in the general situation and, in
that sense, is somewhat misleading.  The reason is that, in
contrast with the three-dimensional Chern-Simons theory, we have
also used the diffeomorphisms to prove the absence of local
degrees of freedom.  Indeed, these diffeomorphisms are needed to
bring $F$ to its canonical form.  But if there are many abelian
fields, then there are many $F$'s to be brought simultaneously
to canonical form and this is not possible with a
diffeomorphism.  Thus, for many ($N>1$) abelian fields, one
expects the existence of local degrees of freedom  unless the
invariant tensor $g_{abc}$ happens to have been chosen in some
peculiar way that enlarges the number of gauge symmetries of the
theory (accidental gauge symmetries).

A typical example of a theory with accidental gauge symmetries
is obtained by taking all the mixed components of $g_{abc}$ to
vanish, so that the action is just the direct sum of $N$ copies
of the action for a single abelian field.  The theory is then
clearly invariant under diffeomorphisms acting independently on
each copy and has no degrees of freedom.  But there is no reason
to take  vanishing mixed components for $g_{abc}$.  If these
mixed components differ from zero (and cannot be brought to zero
by a change of basis), then the action is not invariant under
diffeomorphisms acting independently on each gauge field
component $A^a$, because the invariance of the cross terms
requires the diffeomorphism parameters for each copy to be
equal, thus gluing all of them together in a single symmetry.

In order to substantiate this discussion and to count precisely
the number of local degrees of freedom, it is best to turn to
the Hamiltonian analysis \cite{HT}.  To that end, we shall
assume that the spacetime manifold $M$ has the topology $\Re
\times \Sigma$, where $\Sigma$ is a four-dimensional manifold.
We then decompose the spacetime gauge field 1-form $A^a$ as
$A^a_\mu dx^\mu=A^a_0dt + A^a_i dx^i$ where the coordinate $t$
runs over $\Re$ and the $x^i$ are coordinates on $\Sigma$.
Although there is no spacetime metric  to give any meaning to
expressions such as timelike or spacelike, we will call time to
the coordinate $t$ and we will say that $\Sigma$ is a spacelike
section as shorthand expressions.

It is easy to see that the Chern-Simons action depends linearly
on the time derivative of $A^a_i$,
\bb
I = \int_\Re \int_\Sigma [l^i_a(A^b_j) \dot A^a_i - A^a_0 K_a ]
,
\label{I2}
\ee
where $K_a$ is given by
\bb
K_a =- g_{abc}\epsilon^{ijkl} F^{b}_{ij} F^{c}_{kl}.
\label{K}
\ee
The explicit form of the function $l^i_a(A^b_j)$ appearing in
Eq. (\ref{I2}) is not needed here but only its ``exterior''
derivative in the space of spatial connections, which reads
\bb
\Omega^{ij}_{ab} \equiv
\frac{\delta l^j_b}{\delta A^a_i}-\frac{\delta l^i_a}{\delta A^b_j}
= -4\epsilon^{ijkl} g_{abc} F^{c}_{kl}.
\label{Omega}
\ee

The equations of motion obtained by varying the action with
respect to $A^a_i$ are given by
\bb
\Omega^{ij}_{ab}\dot A^b_j=\Omega^{ij}_{ab}D_jA^b_0 ,
\label{Eq}
\ee
while the variation of the action with respect to $A^a_0$ yields
the constraint $K_a=0$.

Since the action is linear in the time derivatives of $A^a_i$,
the canonically conjugate momenta $p^i_a$ are subject to the $4
N$ primary constraints,
\bb
\phi^i_a = p^i_a - l^i_a \approx 0 \ .
\ee
These constraints transform in the coadjoint representation of
the Lie algebra because the inhomogeneous terms in the
transformation laws of $p_a^i$ and $l_a^i$ cancel out.

It turns out to be more convenient to replace the
constraints $K_a$ by the equivalent set
\bb
G_a=K_a-D_i\phi^i_a.
\ee
The surface defined by $K_a=0$, $\phi^a_i =0$ is equivalent to
the surface defined by $G_a=0$, $ \phi^a_i =0$.  The new
constraints $G_a$ generate the gauge transformations
(\ref{gauge}), e.g. $\{A_i^a,\int_\Sigma\lambda^b
G_b\}=D_i\lambda^a$.

The Hamiltonian action  takes the form \cite{HT},
\bb
I= \int_\Re \int_\Sigma [ p^i_a \dot A^a_i - A^a_0 G_a - u^a_i
\phi^i_a ].
\ee
The Poisson bracket among the constraints is given by
\begin{eqnarray}
\{ \phi^i_a , \phi^j_b \} & =& \Omega^{ij}_{ab}, \\
\{\phi^i_a, G_b\} &=&f^c_{ab} \phi^i_c ,\label{phi-g}\\
\{ G_a,G_b\} &=& f^c_{ab} G_c ,\label{g-g}
\end{eqnarray}
where $f^c_{ab}$ are the structure constants of the Lie algebra,
which vanish in the abelian case that we are considering now.
It follows from the constraint algebra that there are no further
constraints. The consistency condition $\dot G_a=0$ is
automatically fulfilled because $G_a$ is first class while the
other consistency equation
$\dot\phi^i_a=\Omega^{ij}_{ab}u_j^b=0$ will just restrict some
of the Lagrange multipliers $u^b_j$.

Equations (\ref{phi-g}) and (\ref{g-g}) reflect that the
constraints $G_a$  are the generators of the gauge
transformations and that the constraints $\phi_a^i$ transform in
the coadjoint representation.  This means, in particular, that
the $G_a$'s are first class, as mentioned above.

The nature of the constraints $\phi_a^i$ is determined by the
eigenvalues of the matrix $\Omega^{ij}_{ab}$. It turns out that
the matrix $\Omega^{ij}_{ab}$ is not invertible on the
constraint surface $K_a=0$.  Indeed, using some simple
combinatorial identities, one can prove that $K_a$ given in
Eq. (\ref{K}) and $\Omega^{ij}_{ab}$ satisfy the  relation,
\bb
\Omega^{ij}_{ab} F^b_{kj} = \delta^i_k K_a.
\label{null-diff}
\ee
This equation shows that, on the constraint surface $K_a=0$, the
matrix $\Omega^{ij}_{ab}$ has $4$ null eigenvectors
$(v_k)^b_j=F^b_{kj},\ (k=1\ldots 2n)$.  The corresponding $4$
first class constraints, namely
\bb
H_i \equiv F^a_{ij} \phi^j_a,
\ee
generate the spatial diffeomorphisms (\ref{diff2}).  They
satisfy the spatial diffeomorphism algebra, up to gauge
transformations. The presence of these constraints is of course
not a surprise because the Chern-Simons action is invariant
under diffeomorphisms for any choice of the invariant tensor
$g_{abc}$.

One could also expect the presence of another first class
constraint, namely, the generator of timelike diffeomorphisms.
However, as we shall see below, this symmetry is not
independent from the other ones and
hence its generator is a combination of
the first class constraints $G_a$ and $H_i$.

We now examine whether the first class constraints $G_a$ and
$H_i$ are independent and constitute a  complete set. This
depends on the properties of the invariant tensor  $g_{abc}$
and, for  a definite choice of $g_{abc}$, it also depends on the
phase space location of the system. This is due to the fact that
the constraint surface of the Chern-Simons
theory is stratified into phase space regions where
the matrix $\Omega_{ab}^{ij}$ has different ranks.

We will say that
an invariant tensor $g_{abc}$ is {\em generic} if and only if it
satisfies the following condition:
There exist solutions $F^a_{ij}$ of the constraints $K_a=0$ such
that
\begin{enumerate}
\item[ $(i)$]  the matrix $F^b_{kj}$ (with $b,j$ as row index and
$k$ as column index) has maximum rank $4$, so that the only
solution of $\xi^kF^b_{kj}=0$ is $\xi^k=0$ and therefore the $4$
null eigenvectors $(v_k)^b_j=F^b_{kj},\ (k=1\ldots 4)$ are
linearly independent;
\item[$(ii)$] the $(4N)\times (4N)$ matrix
$\Omega_{ab}^{ij}$ has the maximum rank compatible
with $(i)$, namely $4N-4$; in other words, it has no other null
eigenvectors besides $(v_k)^b_j=F^b_{kj},\ (k=1\ldots 4)$.
\end{enumerate}

We will also say that  the solutions
$F^a_{ij}$ of the constraints $K_a=0$ that allow for this
condition to be satisfied are {\em generic}. The reason for this
name comes from the following observation. For a given generic
$g_{abc}$, a solution fulfilling both conditions $(i)$ and
$(ii)$ will still fullfill them upon small perturbations since
maximum rank conditions correspond to inequalities and define
open regions.  Conversely, a solution not fulfilling conditions
$(i)$ or  $(ii)$, i.e., located on the surface where lower ranks
are achieved (defined by equations expressing that some non
trivial determinants vanish), will fail to remain on that
surface upon generic perturbations consistent with the
constraints. Non-generic
solutions of the constraint equations are also of physical
interest but will not be considered here (see Ref.
\cite{BGH2} for a more complete analysis).

The genericallity condition represents the general case in the
sense that it defines an open region in the space of the
invariant tensors.  Indeed, as we have pointed out, these
algebraic conditions enforce  inequalities.  Therefore, to
achieve a lower rank, some extra conditions would have to be
fulfilled and this would put $g_{abc}$ on a surface of lower
dimensionality in the space of the invariant tensors.

The physical meaning of the above algebraic conditions is
straightforward.  They simply express that the gauge transformations
(\ref{gauge}) and the spatial diffeomorphisms (\ref{diff2}) are
independent  and that there are no other first class constraints
among the $\phi^j_a$ besides $H_i$.

In order to illustrate these points and to show that the
genericallity
condition is not self-contradictory and can be actually
fulfilled, let us work out a simple example.  Take a
non-diagonal $g_{abc}$ of the form:
\bb
g_{a11}=0, \hspace{15mm} g_{a'b'1}\equiv g_{a'b'}
\mbox{~~~invertible}
\label{g1111}
\ee
where $a',b',\ldots=2,3\ldots N$. Then, the constraints $K_a=0$
are solved by taking $F^{a'}_{ij}=0$ and $F^1_{ij}$ arbitrary.
The matrix $\Omega^{ij}_{ab}$ has the tensor product form
\bb
\Omega^{ij}_{1a}=0,
\hspace{15mm}
\Omega^{ij}_{a'b'}= g_{a'b'}\epsilon^{ijkl} F^1_{kl}
\ee
and is thus of rank $4(N-1)$ provided that $F^1_{ij}$ is taken
to be invertible. The invertibility of $F^1_{ij}$ also ensures
that the only solution of $\xi^kF^b_{kj}=0$ is $\xi^k=0$.
Therefore, we can conclude that the invariant tensor $g_{abc}$
given in Eq. (\ref{g1111}) is generic.  Also, this example shows
the stratification of phase space.  While the solution
that we have discussed (with $\det (F^1_{ij})\neq 0$) is
generic, solutions of the same form but with $\det (F^1_{ij})=0$
belong to one of these lower dimensional non-generic phase space
regions.

Thus, for generic theories, the only first class constraints are
$G_a=0$ and $H_i=0$, which shows that the generator of timelike
diffeomorphisms is not independent from $G_a$ and $H_i$.
This may be verified explicitly by
writing  the action of a timelike diffeomorphism
parametrised by $\xi^\mu=(\xi^0,0)$ on $A^a_i$ as, see Eq.
(\ref{diff2}),
\bb
\delta_\xi A^a_i=\xi^0F_{i0}^a.
\label{tdiff}
\ee
Now, the equations of motion (\ref{Eq}) are
$\Omega^{ij}_{ab}F^b_{0j}=0$. Since the only zero eigenvectors
of the matrix $\Omega^{ij}_{ab}$ are $F^b_{kj}$, there must
exist some $\zeta^k$ such that $F^b_{j0}=\zeta^k F^b_{jk}$.
Inserting this result in Eq. (\ref{tdiff}), we obtain
\begin{equation}
\delta_\xi A^a_i= \xi^0\zeta^kF_{ik}^a,
\end{equation}
which is an improved spatial diffeomorphism with parameter
$\xi^0\zeta^k$.

We can now count the number of local degrees of freedom in the
generic case.  We have, $2\times 4N$ canonical variables
($A^a_i, p^i_a$), $N$ first class constraints $G_a$ associated
to the gauge invariance, $4$ first class constraints $H_i$
associated to the (spatial) diffeomorphism invariance, and
$4N-4$ second class constraints (the remaining $\phi^i_a$).
Hence, we have
\bb
\frac{1}{2}[8N - 2(N + 4) - (4N - 4)] = 2N-2-N
\label{DF}
\ee
local degrees of freedom.   The formula does not apply to $N=1$
because the spatial diffeomorphisms are not independent in that
case, as can be checked directly on the canonical form $F = dx^1
\w dx^2$. From (\ref{DF}) we see that, for $N=2$, there are no
degrees of freedom. This happens  because one does not use all
the diffeomorphism invariance to bring the first $F^1$ to a
canonical form.  One may then use the residual diffeomorphism
invariance to bring the second field strength $F^2$  to a
canonical form also.  However, for $N>2$, there are degrees of
freedom.

The analysis has been performed so far in the abelian case.  In
the non abelian case, the analysis proceeds similarly, but the
invariance condition strongly restricts the possible  $g_{abc}$.
So one may fear that there could be a conflict between the
invariance condition and the genericallity condition.  This
is not the case and we have verified explicitly that the
three-index invariant tensor of $SU(p)$ ($2<p\leq 6$) is
generic.  Likewise the gravitational Chern-simons theories in
$5$ dimensions are also generic and therefore do carry local
degrees of freedom (this was anticipated in quite a different
way by
Chamseddine who analysed perturbations around a non trivial
background \cite{Chamseddine}).

What has been done in $5$ dimensions can be repeated in higher
(odd) dimensions.  Provided the invariant tensor $g_{a_1 \dots
a_{n+1}}$ fulfills a genericallity condition
that is the straightforward generalization of the one
appropriate to $5$ dimensions, one finds that the canonical
formulation of Chern-Simons theory involves $N+2n$ first class
constraints and $2nN-2n$ second class constraints in the generic
case.  The first class constraints generate the gauge symmetries
(\ref{gauge}) and the spatial diffeomorphisms (\ref{diff2}).  As
in $5$ dimensions, the timelike diffeomorphisms can be expressed
in terms of the other gauge symmetries.  Since there are $2nN$
conjugate pairs, the number of local degrees of freedom is equal to
\bb
\frac{1}{2}[4nN - 2(N + 2n) - (2nN - 2n)] = nN-n-N,
\label{dfree}
\ee
where $N>1$ and $n>1$.  This expression
vanishes only for $n=2$ and $N=2$.  Again, one may also verify
that the genericallity condition is not self-contradictory by
exhibiting invariant tensors that fulfill it.  For instance, one
may take a direct generalization of Eq. (\ref{g1111}).  The complete
analysis, where the explicit isolation of the second class
constraints is performed and the Dirac bracket is computed, will
be reported elsewhere \cite{BGH2}.

When the invariant tensor $g_{a_1\ldots a_{n+1}}$ is not
generic, $\Omega_{ab}^{ij}$ has further zero eigenvalues   and
thus, there are further gauge symmetries.  This implies that the
number of degrees of freedom is smaller than in the generic case
and may even vanish. As we mentioned above, an extreme example
is given by $N$ uncoupled abelian gauge fields, where the extra
gauge symmetries are diffeomorphisms acting independently on
each individual copy.

To conclude, we have shown that higher dimensional Chern-Simons
theories, even though constructed along the same topological
pattern as in  $2+1$ dimensions, do have local degrees of
freedom provided that the invariant tensor that enters the
action fulfills an appropriate genericallity condition.  This
condition implies that there are no accidental gauge symmetries.
The result cannot be  anticipated by analysing the case of a
single abelian field, which is not representative of the general
case.

{\bf Acknowledgments.} We are grateful to Claudio Teitelboim and
Jorge Zanelli for useful conversations.  M.B. is partially
supported by grants 1930910-93 and 1940203-94 from FONDECYT
(Chile), a grant from Fundaci\'on Andes (Chile), and by
institutional support to the Centro de Estudios Cient\'{\i}ficos
de Santiago provided by SAREC (Sweden) and a group of Chilean
private companies (COPEC,CMPC,ENERSIS).  L.J.G. is supported by
a joint fellowship from the Ministerio de Educaci\'on y Ciencia
(Spain) and the British Council.  This work has been supported
in part by research funds form the Euopean Community and by a
research grant form F.N.R.S.


\begin{references}

\bibitem{Jackiw} R. Jackiw, {\it Phys. Rev. Lett} {\bf 41},
1635 (1978)

\bibitem{Chamseddine} A. H. Chamseddine, {\it Nucl. Phys.}
{\bf 346}, 213 (1990);
{\it Phys. Lett.} {\bf B233}, 291 (1989)

\bibitem{at86} A. Ach\'ucarro, P. Townsend, {\it
Phys. Lett.} {\bf B180}, 89 (1986).
E. Witten, {\it Nucl. Phys.} {\bf B311}, 46 (1998)

\bibitem{Lovelock} D. Lovelock, {\it J. Math. Phys.} {\bf 12},
498 (1971)


\bibitem{BTZ1} M. Ba\~nados, C. Teitelboim, J. Zanelli,
{\it Phys. Rev.} {\bf D49}, 975 (1994)

\bibitem{BTZ2} M. Ba\~nados, C. Teitelboim, J. Zanelli,
{\it Phys. Rev. Lett.} {\bf 69}, 1849 (1992).
M. Ba\~nados, M. Henneaux, C. Teitelboim,
J. Zanelli, {\it Phys. Rev.} {\bf D48}, 1506 (1993)

\bibitem{fp89}  R. Floreanini, R. Percacci,  {\it Phys.
Lett.} {\bf B224}, 291 (1989).

\bibitem{HT} For a recent treatment on the Dirac method see,
M.  Henneaux and C. Teitelboim, {\em Quantization of Gauge
Systems} (Princeton University Press, Princeton, 1992).

\bibitem{BGH2} M. Ba\~nados, L. J. Garay, M. Henneaux,
{\it The dynamical structure of higher dimensional
pure Chern-Simons theories}, in
preparation.

\end{references}
\end{document}